# 3D quench modeling based on T-A formulation for high temperature superconductor CORC cables

Yawei Wang[1], Jinxing Zheng[2], Zixuan Zhu[1], Min Zhang[1] [*], Weijia Yuan[1] [*]

[1]Department of Electronic and Electrical Engineering, University of Strathclyde, G1 1XQ, Glasgow, UK.

[2]Institute of Plasma Physics, Chinese Academy of Science, 230031, Anhui, P. R. of China.

E-mail: min.zhang@strath.ac.uk; weijia.yuan@strath.ac.uk

**Abstract**

High temperature superconductor (HTS) (RE)Ba$_2$Cu$_3$O$_x$ (REBCO) conductor on round core cable (CORC) has high current carrying capacity for high field magnet and power applications. In REBCO CORC cables, current redistribution occurs among tapes through terminal contact resistances when a local quench occurs. Therefore, the quench behaviour of CORC cable is different from single tape situation, for it is significantly affected by terminal contact resistances. To better understand the underlying physical process of local quenches in CORC cables, a new 3D multi-physics modelling tool for CORC cables is developed and presented in this paper. In this model, the REBCO tape is treated as a thin shell without thickness, and four models are coupled: *T*-formulation model, A-formulation model, a heat transfer model and an equivalent circuit model. The current redistribution, temperature and tape voltage of CORC cable during hot spot induced quenches are analysed using this model. The results show that the thermal stability of CORC cable can be considerably improved by reducing terminal contact resistance. The minimum quench energy (MQE) increases rapidly with the reduction of terminal contact resistance when the resistance is in a middle range, which is about $5\ \mu\Omega \leq R_t \leq 200\ \mu\Omega$ in this study. When the terminal contact resistance is too low or too high, the MQE shows no obvious variation with terminal contact resistances. With a low terminal contact resistance, a hot spot in one tape may induce an over-current quench on the other tapes without hot spots. This will not happen in a cable with high terminal contact resistance. In this case, the tape with hot spot will quench and burn out before inducing a quench on other tapes. The modelling tool developed can be used to design CORC cables with improved thermal stability.

## 1. Introduction

The second generation (2G) high temperature superconductor (HTS), $(RE)Ba_2Cu_3O_x$ (REBCO) coated conductor, shows great potentials in many applications, owing to its high current density and high operational temperatures[1]. Some applications, like high field magnet and power transmission, require large current capacity[2-5]. Thus, three cabling concepts have been developed: Roebel cable[6-10], twisted stack cable[11-14], and conductor on round core (CORC) cable[4, 5]. In CORC cables, the HTS tapes are wound helically on a round former, thus, the cable has great advantages on flexibility, mechanical strength, high current density & capacity, and AC loss reduction[4, 15-19]. Tests have been performed on CORC cables in background fields up to 20 T, and a HTS magnet wound with CORC cables eventually reached a magnetic field of up to 20 T [20-23].

It has been reported that REBCO conductor may be damaged by temperature rise during a local quench [24-27]. Therefore, quench becomes one of the largest challenges for HTS applications. The quench behaviour of single 2G HTS tape has been studied by both experiments and simulations[28-31]. But the quench behaviour of CORC cables is different from single HTS tapes. In CORC cables, HTS tapes are soldered to copper leads, the variation of terminal contact resistance often leads to an inhomogeneous current distribution among parallel tapes [32]. During a local quench, transport current will be redistributed among these parallel tapes through terminal contact resistances. Therefore, the quench behaviour of CORC cable is closely related to the terminal contact resistance. There are very little studies on 2G HTS CORC cable quenches [33], due to the difficulty of measuring current distribution within the cable. Therefore, a modelling tool is essential to help understand its quench behaviour under the influence terminal contact resistance. So far, however, most of present modelling study on HTS focuses on its electromagnetic behaviours [34-40]. There is no multi-physics quench modelling for 2G HTS CORC cables.

This paper presents a modelling study on the quench behaviour of HTS CORC cables. A novel multi-physics 3D quench model is developed based on commercial finite element method (FEM) software. This model is based on the T-A formulation and coupled with a heat transfer model and an equivalent circuit model in COMSOL Multiphysics. The current redistribution, temperature rise, normal zone propagation in the CORC cable are analysed during a hot-spot induced quench. The effect of terminal contact resistance on the local quench process is studied. The variation of minimum quench energy due to terminal contact resistance is discussed. This model enables the study of CORC quench behaviours to be practical, simple and extendable.

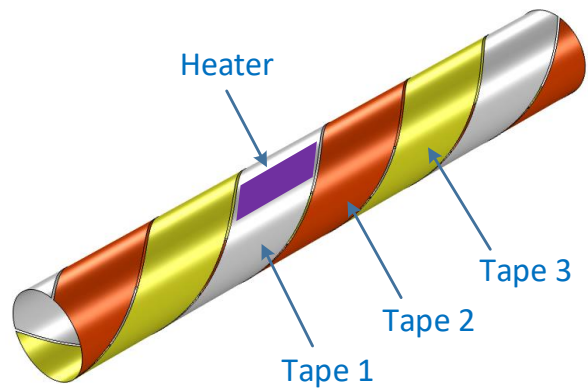

**Figure 1.** The geometry of the single-layer CORC cable consisting of three REBCO tapes.

## 2. HTS cable model studied

A HTS CORC cable sample is built for the quench study in this paper, which is based on the cable model in reference [17, 41, 42]. This cable has a single layer with three REBCO tapes, as shown in Figure 1. The angular of the cable is 40°. The ReBCO tape used here has a width of 4 mm and a thickness of 0.1 mm. The critical current of the tape is 100 A in self-field at 77 K. The operation temperature is 70 K, at which the tape's critical current is about 148 A. The transport current of each tape is 120 A, so that 360 A current in total is delivered by this cable. A heater is added on one tape to induce a local hot spot. The tape zone covered by this heater is 26.11 mm$^2$. The tape with heater is labelled as Tape 1, and the other two tapes are labelled as Tape 2 and Tape 3 respectively, as shown in Figure 1. More specifications about this cable is shown in the Table 1.

**Table 1.** Specification of the CORC cable model

| | Parameters | Quantity |
|---|---|---|
| Cable | Inner diameter | 5.2 mm |
| | Angular | 40° |
| | Pitch | 19 mm |
| | Operating current | 360 A |
| | Operating temperature | 70 K |
| | Tapes per layer | 3 |
| REBCO tape | Width/Thickness | 4 mm/0.1mm |
| | Critical current, $I_c$ @ 77K/70K | 100 A/ 148 A |
| | Critical temperature | 92 K |
| | Operating current | 120 A |
| | Substrate (Hastelloy) | 50 $\mu m$ |
| | Stabilizer (Copper) | 2*20 $\mu m$ |

## 3. Numerical quench model

The 3D quench model consists of four modules, T-formulation model, A-formulation model, thermal model and equivalent circuit model, as shown in Figure 2. A thin shell assumption is applied on the coated conductor, thickness of the REBCO conductor is neglected, while its geometry structure along width and length direction remains the same, as shown in Figure 3. The T-formulation is applied to the conductor shell to calculate current distribution. The A-formulation is applied to the whole 3D domain to calculate magnetic field induced. The magnetic field obtained in A-formulation model is fed back to the T-formulation model, so that the electromagnetic coupling is achieved. The thermal model is applied to the conductor shell only, which is to calculate the temperature distribution. The heat source includes the external heat pulse and Joule heat generated by current. Temperature is fed back to the T-formulation model for the calculation of critical current. The equivalent circuit model is to calculate the current redistribution among tapes.

All these models are built and solved in COMSOL. The T-formulation model is solved in a PDE package, the A-formulation model is solved in magnetic field package, the thermal model is solved in a heat transfer on shell package, and the circuit model is solved by an ordinary differential equation (ODE) package.

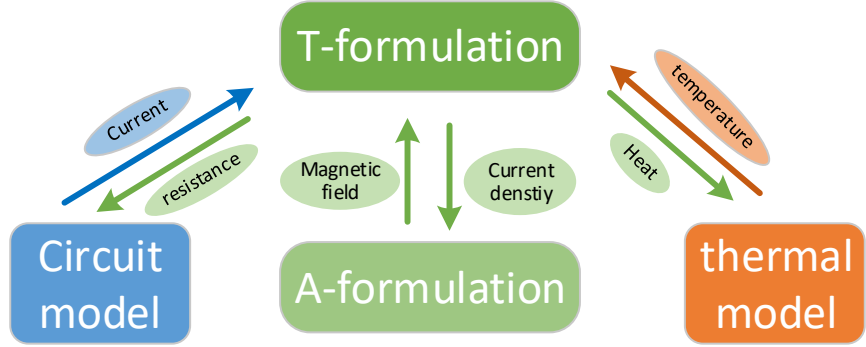

**Figure 2.** The numerical algorithm of the 3D quench model for CORC cables.

*3.1. Thin shell approximation*

The REBCO coated conductor is treated as a thin shell to effectively reduce number of meshes as well as computation cost. It is reasonable to neglect the variation of magnetic field and temperature along the thickness direction, due to the high aspect ratio of REBC conductor [34, 43]. However, current sharing occurs between superconducting layer and metallic stabilizer during quench operation, which has to be considered for the quench modelling.

As shown in Figure 3(a), the superconducting layer is covered by metallic layers (substrate and stabilizer). Transport current flows in the superconducting layers below critical current since the resistivity of the stabilizer layer is much higher than that of superconductor. The E-J relationship can be expressed as:

$$\mathbf{E}(\mathbf{J}) = E_0 \left( \frac{|\mathbf{J}|}{J_c(\mathbf{B}, T_h)} \right)^n \frac{\mathbf{J}}{J_c(\mathbf{B}, T_h)} \qquad (1)$$

where $E_0 = 1\times10^{-4}$ V/m, $J_c$ is the critical current density, depending on the background field $\mathbf{B}$ and temperature $T_h$. During a local quench, the resistance of the superconducting layer in normal zone increases dramatically, and some currents are forced out to flow into metallic layers. In this situation, the REBCO conductor is equivalent to two parallel resistances: the resistance of superconducting layers $R_{su}$ and that of the metallic layers $R_m$, as shown in the Figure 2(b) [44]. The governing equation of this equivalent circuit can be derived from Kirchhoff's law:

$$\begin{cases} E_0 \left( \dfrac{I_{su}}{I_c(B,T_h)} \right)^n - I_m \rho_m(T_h) = 0 \\ I = I_{su} + I_m \end{cases} \quad (2)$$

where $I$ is the total current flowing in the REBCO conductor, $I_{su}$ is the current flowing in superconducting layer, and $I_m$ is the current flowing in metallic layers. $\rho_m$ is the resistivity of the metallic layers, which is practically equal to the resistivity of the copper stabilizer [44]. It can be calculated using the following formula:

$$\rho_m(T_h) = \rho_{copper}(300)[1 + 0.0039(T_h - 300)] \quad (3)$$

Therefore, the *E-J* power law of REBCO conductor considering the overcurrent operation is revised as:

$$\begin{cases} E(J) = (J - J_{su})/\sigma_m(T_h) \\ J = \dfrac{I}{S_c}; \ J_{su} = \dfrac{I_{su}}{S_c}; \ J_m = \dfrac{I_m}{S_c} \end{cases} \quad (4)$$

where $S_c$ is the area of the conductor's cross section, $J$ is the engineering current density flowing through the tape, $J_{sc}$ and $J_m$ are the equivalent current density flowing in superconducting layer and metallic layers respectively.

The dependence of the critical current on the temperature and external magnetic field can be expressed as [30, 45-47]:

$$\mathbf{J}_c(\mathbf{B}, T_h) = \dfrac{I_{co}}{S_c} \cdot J_{cT}(T_h) \cdot J_{cB}(B_{par}, B_{per}) \quad (5)$$

$$J_{cT}(T_h) = \begin{cases} \left( \dfrac{T_c - T_h}{T_c - T_o} \right)^\beta & \text{if } T_h < T_c \\ 0 & \text{if } T_h \geq T_c \end{cases} \quad (6)$$

$$J_{cB}(B, \theta) = J_{cB}(B_{par}, B_{per}) = \dfrac{1}{[1 + \sqrt{(kB_{par})^2 + B_{per}^2}/B_c]^b} \quad (7)$$

where $\beta = 1 \sim 2$ represents the temperature dependence of critical current, $T_o = 77$ K, and $T_c = 92$ K is the critical temperature of REBCO conductor. $B_{par}$ and $B_{per}$ represent the magnetic fields parallel and perpendicular to the tape surface respectively, which will be obtained from the A-formulation model. $k$, $b$, $B_c$ are shape parameters representing the field anisotropic characteristics of REBCO conductors, which are $k = 0.0605$, $b = 0.7580$, and $B_c = 103$ mT in this study. $I_{co}$ is the critical current at 77 K in self-field, which is 100 A in the following study.

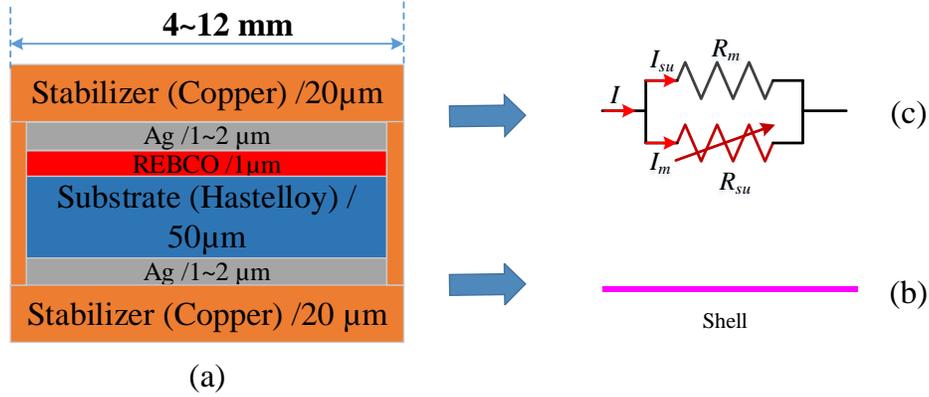

**Figure 3.** (a) Cross section of the REBCO conductor used in this study, which is from Superpower; (b) The approximated thin shell of the conductor; (c) the equivalent circuit model for the REBCO conductor.

*3.2. T- formulation*

The *T*-formulation model is applied to the REBCO conductor shell only. Based on the thin shell approximation, the current component normal to the tape surface is neglected, and the current only flows along the tangential direction of the tape surface. Therefore, a scalar variable *T* is defined to express the current density on this thin shell conductor[43, 48, 49]:

$$\mathbf{J} = \nabla \times (T\mathbf{n}) \qquad (8)$$

where *T* is the normal component of the current vector potential **T**, and it is applied to the REBCO conductor shell only. $\mathbf{n} = [n_x, n_y, n_z]$ is the unit normal vector of the tape surface.

The governing equation of the model is derived from Faraday's law using variable *T*:

$$\nabla \times \mathbf{E}(\mathbf{J})) = -\frac{\partial \mathbf{B}}{\partial t} \qquad (9)$$

where the magnetic flux density **B** is obtained directly from the A-formulation model. Notice that the magnetic field **B** here is not solved in the iteration of Equation (9), and it has no direct coupling with variable $T$. Therefore, a vector problem with three variables in 3D space is reduced to a problem of the scalar variable $T$ in a 2D thin shell, which significantly reduces the computation cost. A boundary condition is added on the edges of the conductor shell to impose the transport current:

$$T_1 - T_2 = \frac{I_k}{d}, \quad (k = 1, 2, 3) \tag{10}$$

where $d$ is the thickness of the REBCO conductor, $T_1$ and $T_2$ are the current vector potential on the two edges of the tape, as shown in Figure 4. $I_k$ is the transport current of the $k$-th tape, which is calculated from the equivalent circuit model in section 2.4.

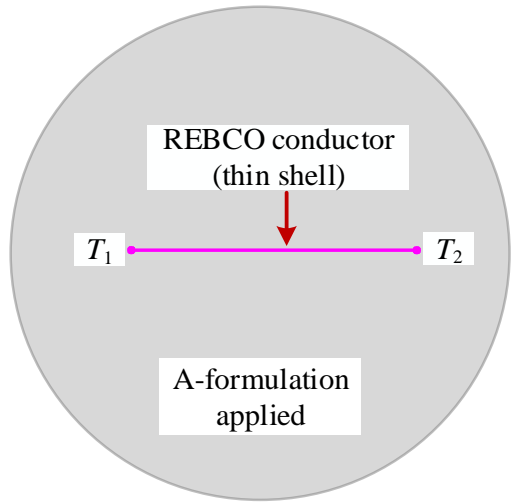

**Figure 4.** Domain relationship between the **A**-formulation model and $T$-formulation model.

*3.3.* **A**-*formulation*

To calculate the magnetic field, the **A**-formulation model is applied to all domains including the shell conductor, as shown in the Figure 3. Here **A** is the magnetic potential, defined as:

$$\mathbf{B} = \nabla \times \mathbf{A} \tag{11}$$

Coupling it to Ampere's law, the governing equation can be expressed as:

$$\nabla \times \nabla \times \mathbf{A} = \mu \mathbf{J} \tag{12}$$

A boundary condition (B.C.) is added on the thin shell conductor to impose the surface current,

$$\mathbf{n} \times (\mathbf{H}_1 - \mathbf{H}_2) = \mathbf{J} \tag{13}$$

where μ is the permearbility, $\mathbf{H}_1$ and $\mathbf{H}_2$ are the magnetic fields on both sides of the conductor shell. The current density $\mathbf{J}$ is obtained from the *T*-formulation model and is applied to the REBCO conductor shell. The magnetic field **B** obtained from this model will be fed back to the T-formulation model.

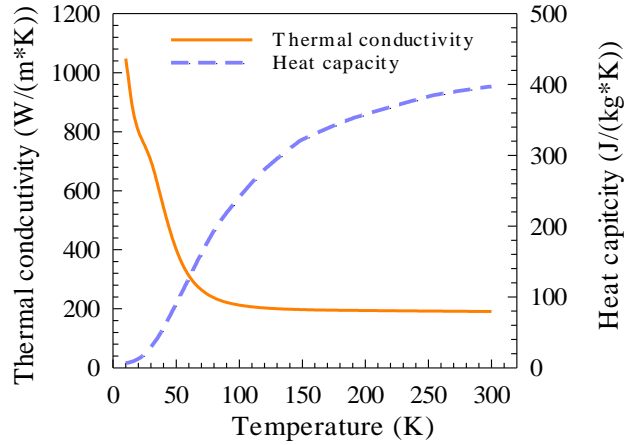

**Figure 5.** The equivalent heat capacity and thermal conductivity of REBCO conductor used in this study.

*3.4. Heat transfer model*

The heat transfer model is to calculate the temperature on REBCO conductors. Given the thin structure of the REBCO conductor and the high thermal conductivity of metallic stabilizer, the temperature difference perpendicular to the conductor surface is neglected, assuming the conductor has a uniform temperature along the thickness direction. Therefore, the REBCO conductor is also approximated to a thin shell in this heat transfer model.

The governing equation of a typical heat transfer model in solid is as follows,

$$\rho C_p \frac{\partial T_h}{\partial t} + \nabla \cdot (-k \nabla T_h) = Q_s + Q_p \tag{14}$$

where $C_p$, $k$ are the heat capacity and thermal conductivity respectively, both of which are temperature dependent. $\rho$ is the density. $Q_s$ is the heat source power generated by the transport current in stabilizer; $Q_p$ is the heat pulse power imposed, which induces a hot spot on the RBECO tape in the simulation. On the thin shell, the gradient operator $\nabla T_h$ can be decomposed to a normal component $\nabla_n T_h$ and a tangential component $\nabla_t T_h$.

$$\begin{cases} \nabla_n T_h = (\nabla T_h \cdot \mathbf{n})\mathbf{n} \\ \nabla_t T_h = \nabla T_h - (\nabla T_h \cdot \mathbf{n})\mathbf{n} \end{cases} \quad (15)$$

where $\mathbf{n}$ is the unit normal vector of the shell. The equation (14) can be rewritten as:

$$\rho C_p \frac{\partial T_h}{\partial t} - k\nabla_n \cdot (\nabla_n T_h) - k\nabla_t \cdot (\nabla_t T_h) = Q_s + Q_p \quad (16)$$

Only the tangential component is considered in the thin shell. Therefore, the governing equation for the conductor shell is as follows [45]:

$$\rho C_p \frac{\partial T_h}{\partial t} - k\nabla_t \cdot (\nabla_t T_h) = \frac{Q_s + Q_p}{d} \quad (17)$$

$$\mathbf{n} \cdot (-k\nabla T_h) = h(T_h - T_e) \quad (18)$$

where is $d$ is the thickness of the REBCO conductor, $T_e$ is the environment temperature, and $h$ is the coefficient of heat transfer for the cooling. Eq. 18 is the cooling boundary condition, which is applied to the shell surface. Cooling on the edges of the conductor shell is neglected. In the study of this paper, an adiabatic boundary condition is applied to the tape, because the quench occurs in a very short time compared to time constant of thermal process.

Equivalent homogenous thermal conductivity $k$ and heat capacity $C_p$ are applied to the thin shell. They are calculated as parallel thermal resistances using following formula:

$$k = \frac{k_1 S_1 + k_2 S_2}{S_1 + S_2} \quad (19)$$

$$C_p = \frac{C_{p1} m_1 + C_{p2} m_2}{m_1 + m_2} \quad (20)$$

where $S_1$ and $S_2$ are the cross section areas of substrate and stabilizer respectively, $k_1$ and $k_2$ are the thermal conductivity of substrate and stabilizer respectively, $C_{p1}$ and $C_{p2}$ are the thermal capacity of substrate and stabilizer respectively; $m_1$ and $m_2$ are the mass of the substrate and stabilizer respectively. The equivalent thermal conductivity and capacity of REBCO tape used in this study are shown in Figure 5.

*3.5. Equivalent circuit model and CORC cable sample*

An equivalent circuit model is developed for this CORC cable to calculate the current redistribution among tapes during a local quench, as shown in Figure 6. $R_{t,k}$ is the terminal soldering resistance of the *k*-th tape, which ranges from 0.1 $\mu\Omega$ to 200 $\mu\Omega$ in previous measurements [32]. $R_{tape,k}$ is total resistance of the *k*-th REBCO tape, including resistances of superconducting layer and stabilizer. The governing equation of this circuit model is as follows:

$$\begin{cases} u_k = I_k(R_{t,k} + R_{tape,k}) + \sum_{n=1}^{3} \frac{dI_n}{dt} M_{k,n}; (k=1,2,3) \\ u_1 = u_2 = u_3 \\ I_{op} = I_1 + I_2 + I_3 \end{cases} \quad (21)$$

where $u_k$ and $I_k$ are the voltage and current of *k*-th tape respectively. $I_k$ is fed back to the T-formulation model. $I_{op}$ is the total current delivered by the cable. $M_{k,n}$ is the mutual inductance between *k*-th tape and *n*-th tape, and it also represents the self-inductance when *k*=*n*. Due to the symmetry of the three tapes in a CORC structure, their mutual inductances and self-inductance $M_{k,n}$ are the same. The current redistribution among three tapes in the same layer of a CORC cable mainly depends on the conductor resistance $R_{tape}$ and terminal contact resistance $R_c$. Here the voltage on terminal contact resistance is called 'contact voltage' $u_{t,k}$. The voltage of the tape does not include the terminal voltage, and it is called 'tape voltage' $u_{tape,k}$.

Current redistribution to metallic cores (copper or stainless steel) is not considered in this study. Firstly, the pressed contact resistance between metallic cores and HTS tapes is much higher than the soldering terminal contact resistance. Experiments on ReBCO tapes have shown that no current is redistributed through pressed contact joints during a local quench. Secondly, in practical applications, the CORC cables often consists of dozens of HTS tapes, the total transport current is below 75% of $I_c$ of the cable. If local quench happens on one tape, the other tapes can absorb all the redistributed current from the quenching tape, meanwhile still operate below critical current or slight overcurrent. Therefore, no current will be generated on the metallic core during a hot spot induced quench, since the resistivity of metallic core is much higher than that of superconductors. If major part of the tapes quench simultaneously, there may be a current redistributed to metallic core.

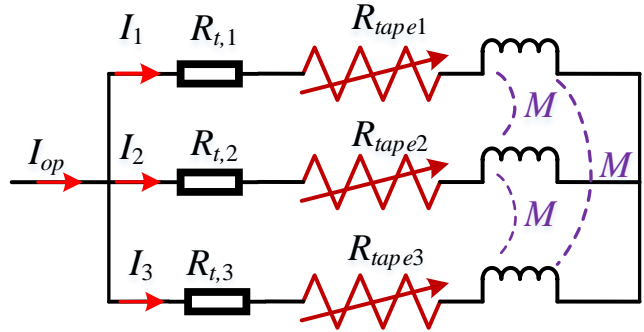

**Figure 6.** The equivalent circuit model for the single-layer CORC cable with three REBCO tapes.

## 4. Results and discussion

### 4.1. Physical process of hot-spot induced quench

In this section, the hot spot induced local quench is studied using above model. The terminal contact resistances of all the tapes, $R_t$, are assumed to be uniform. The impact of non-uniform terminal contact resistances will be presented in future paper. The simulation process is as follows: firstly, the transport current of the CORC cable is ramped to 360 A, and the transport current of each tape is 120 A. Then, a heat pulse is imposed on Tape 1 to induce a hot spot. The period of the heat pulse is 100 ms and the heating power is constant during the heat pulse.

Figure 7 shows the depictions of the temperature and current density during a hot spot induced quench with heat pulse 144 mJ and terminal contact resistance 100 $\mu\Omega$. In figure 7(c), the top temperature of the color legend for each temperature distribution is limited to $T_c$=92 K to highlight the normal zone. The maximum temperature of the heating zone is pointed out by text. Figure 7(a) is the current flowing in superconducting layers. Figure 7(b) is the current flowing in metallic (copper) layers. The results show that some of transport current in superconducting layer is forced out to copper layers when temperature rise is generated by heat pulse. Before that all the transport current flows in superconducting layer bellow critical current. After the heater is turned off, current in copper layer continues generating heat. If the local heat accumulation is dissipated timely to other zones of the tape, the conductor will recover automatically from this heat pulse. In the case of Figure 7, the local heat accumulation is faster than the heat dissipation along the tape, temperature increase continuously and a quench is induced after heat pulse.

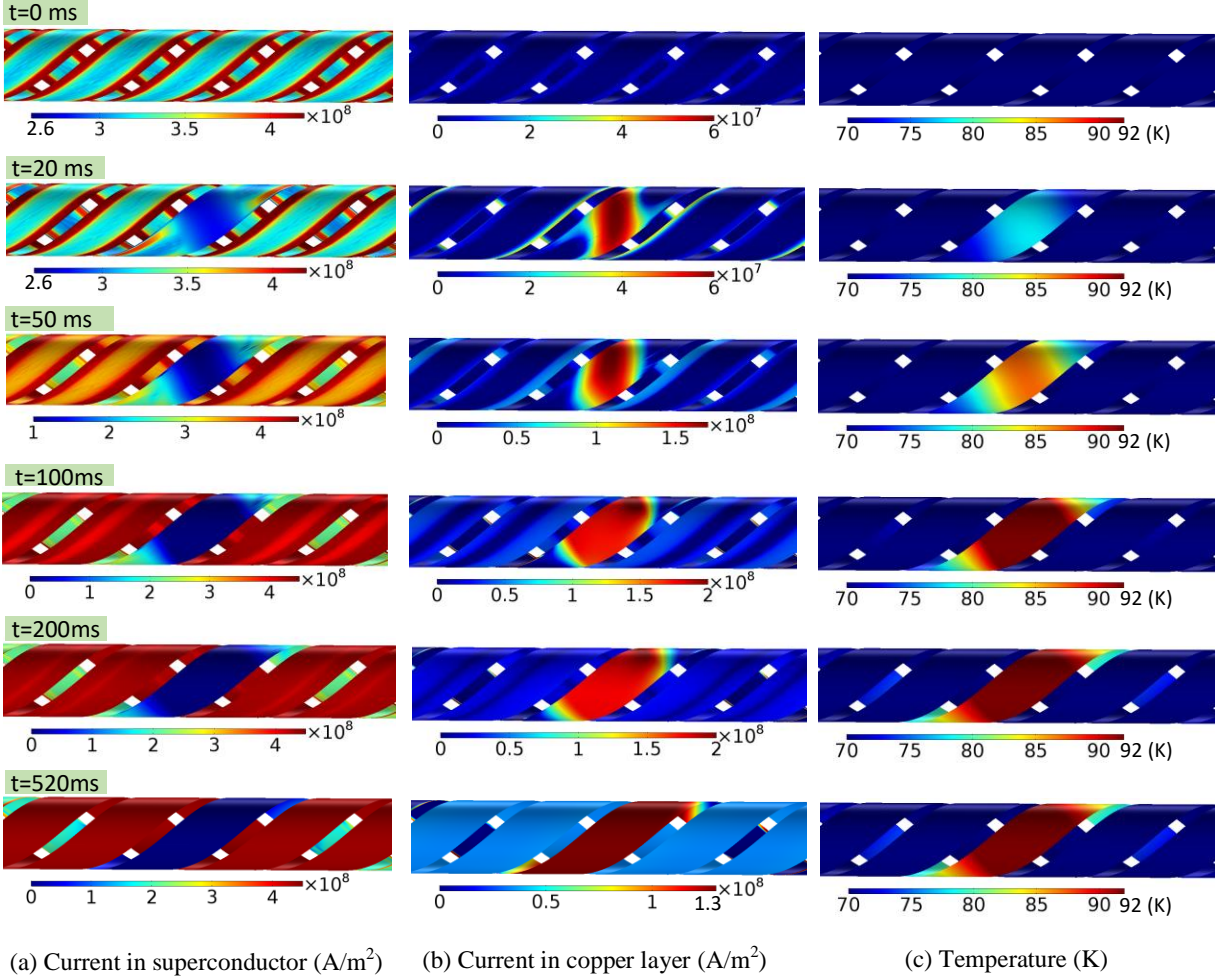

(a) Current in superconductor (A/m$^2$)   (b) Current in copper layer (A/m$^2$)   (c) Temperature (K)

**Figure 7.** Distributions of current density in superconducting layer, current density in metallic layers and the temperature during a hot-spot induced quench, heat disturbance 183 mJ, terminal contact resistance 100 $\mu\Omega$.

*4.2. Influence of terminal contact resistance*

This section is study the influence of terminal contact resistance Rt on the quench process of HTS CORC cables. The initial operating current is still 120 A for each tape, and the period of heat pulse is 100 ms. A series of simulations are conducted under different terminal contact resistances and heat pulses. The results show the thermal stability of CORC cables is much higher than the single tape situation, because current of the tape with hot spot can be forced out to other tapes through the terminal contact resistance. Reducing the terminal contact resistance can significantly enhance the thermal stability of the cable. When the terminal contact resistance is high, the current redistribution mainly depends on the terminal contact resistance. When the terminal contact resistance is low enough, the current redistribution will be limited by the critical current of tapes without hot spot.

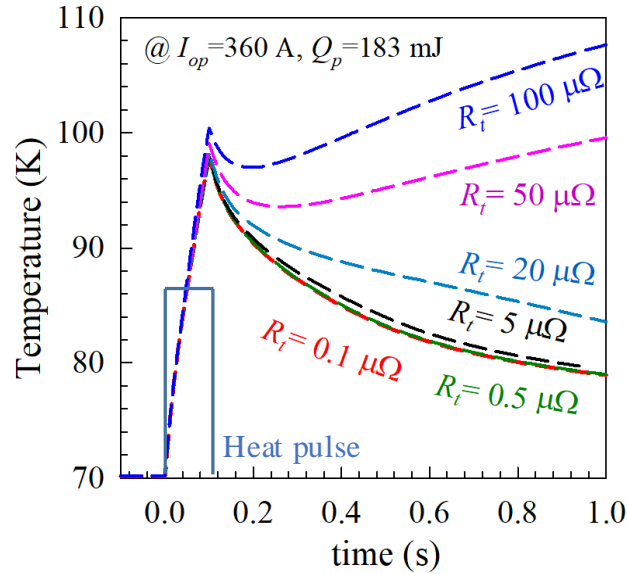

**Figure 8.** Temperature of the hot spot under different terminal contact resistances, heat pulse 183 mJ.

Figure 8 shows the temperature of hot spot under different terminal contact resistances $R_t$ when a heat pulse of 183 mJ is imposed. The temperature increases rapidly when the heat pulse is applied, and reaches to a peak value at the end of the heat pulse ($t$=100 ms). A normal zone is induced by the hot spot and propagates to near zones. When the terminal resistance is low ($R_t \leq$ 20 $\mu\Omega$), the temperature of hot spot drops continuously after the heater is turned off, and finally the hot spot has a recovery from normal zone. Notice that when the terminal contact resistance is low enough ($R_t \leq 5\ \mu\Omega$), the temperature of hot spot shows no obvious difference under different resistances. When the terminal resistance is high ($R_t \geq 50\ \mu\Omega$), the temperature of hot spot shows a slight decrease after the heater is turned off, then increases continuously, and a quench is induced eventually.

To understand this result, we plot in Figure 9 the variation of the transport current and tape voltage in the process. Notice that the tape voltage $u_{tape}$ doesn't include the voltage of terminal contact resistance, it only includes the voltage of REBCO tapes. The results show that transport current is redistributed among tapes through terminal contact resistances when a hot spot is generated, due to the increases of tape voltage. The whole process can be divided into two stages: Stage I, $t$=0~100 ms, the heat pulse is applied; Stage II, $t$ >100 ms, the heat pulse is turned off.

During stage I, $t$=0~100 ms, the current of Tape 1 with hot spot drops continuously due to the temperature rise of hot spot and normal zone propagation. The redistributed current is absorbed by other tapes. This current redistribution is mainly limited by two factors: terminal resistance and critical current. When the terminal resistance is low ($R_t \leq 5\ \mu\Omega$), the critical current seems to play a more important role. As shown in Figure 8($a_1$) and ($a_2$), from 0 ms to 28 ms, Tape 2&3 is below critical current 148 A, the tape resistance is nearly zero, so that a slight voltage rise can lead to a significant current redistribution. Thus, there is no significant voltage rise on the tapes. After $t$=28 ms, Tape 2&3 flows an over-current, and their tape voltages begin to increase rapidly. This voltage rise prevents further redistributed currents from Tape 1. Thus, current in Tape 1 drops much slower than that in previous period $t$<28 ms. At $t$=100 ms, the temperature of hot spot rises to about 100 K, and a considerable normal zone is generated in Tape 1, as shown in Figure 9(a). However, the transport current of Tape 1 only drops to 40 A, and Tape 2&3 flow an over-current 160 A, Tape 1 cannot force out more current to Tape 2&3.

The voltage of Tape 1 with hot spot is approximately same with that of Tape 2&3 when the terminal contact resistance is low enough ($R_t \leq 5\ \mu\Omega$). As shown in Figure 9($a_2$), at $t$=100 ms, the tape voltages of Tape 1 and Tape 2&3 are about 6 mV, which are generated by normal zone in Tape 1 and over-current in Tape 2&3. Meanwhile, the currents of Tape 1 and Tape 2&3 are 40 A and 160 A respectively; the contact voltage of Tape 1 is 4 $\mu$V, the contact voltage of Tape 2&3 is 16 $\mu$V. The voltages of terminal contact resistances are negligible in comparison to the tape voltage of 6 mV. With the increase of terminal contact resistance, the contact voltage takes more and more proportion in comparison to the tape voltage. Obvious difference occurs on tape voltage when the terminal contact resistance is higher than 5 $\mu\Omega$. Therefore, terminal voltage resistance plays a more important role on the current redistribution when the terminal contact resistance is high ($R_t > 5\ \mu\Omega$). With a terminal resistance of 100 $\mu\Omega$, the tape voltage of Tape 1 is 12 mV at $t$=100 ms, and the contact voltage of $R_t$ is 6 mV at the same time. Thus, the lowest current of Tape 1 is only about 60 A (at $t$=100 ms), though a higher temperature is generated on hot spot and a larger normal zone is induced, as shown in Figure 10(b). High terminal resistance prevents current redistributing among tapes, which generates more heat and higher temperature. Therefore, higher terminal resistance leads to higher risk of quench.

During stage II, $t >$100 ms, the cable recovers from hot spot when the terminal contact resistance is low enough, which is $R_t \leq 20\ \mu\Omega$ in this study. The current and tape voltage of all tapes gradually recover to the initial state. The normal zone starts to shrink and finally disappears, as shown in Figure 10(a), since extra heat would be pulled out by cooling system in practical applications. When the terminal contact resistance is high ($R_t > 50\ \mu\Omega$), a quench is finally induced on Tape 1. The tape voltage increases fast with the normal zone propagation. The current of Tape 1 drops continuously, and the redistributed current leads to a continuous current increase on other tapes, as shown in Figure 9($c_1$). This over-current may finally leads to an over-current quench on Tape 2&3, if the protection system does not start timely.

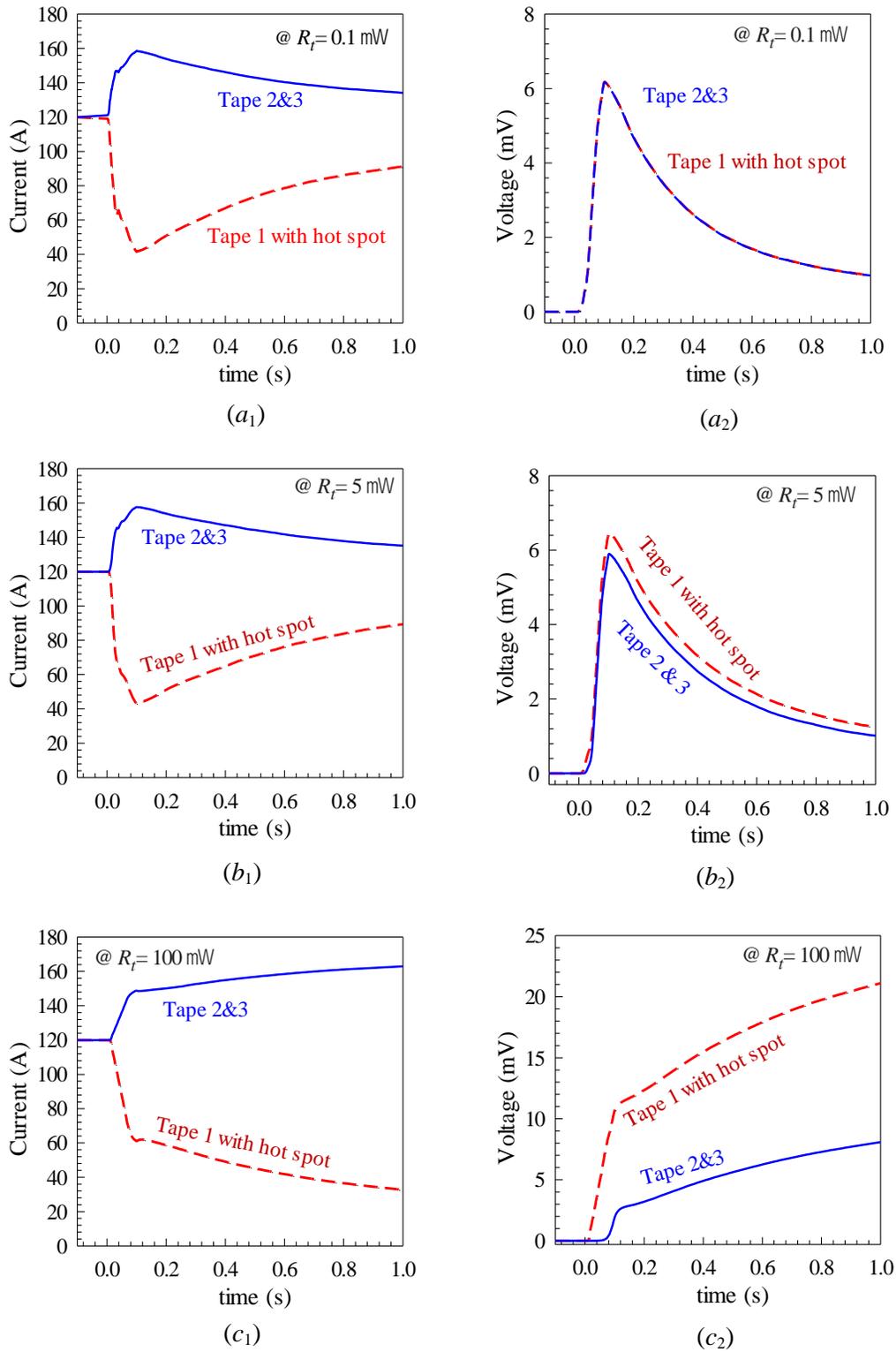

**Figure 9.** Current and voltage of each tape under different terminal contact resistances; total transport current is 360 A, a heat pulse of 183 mJ is applied to Tape 1.

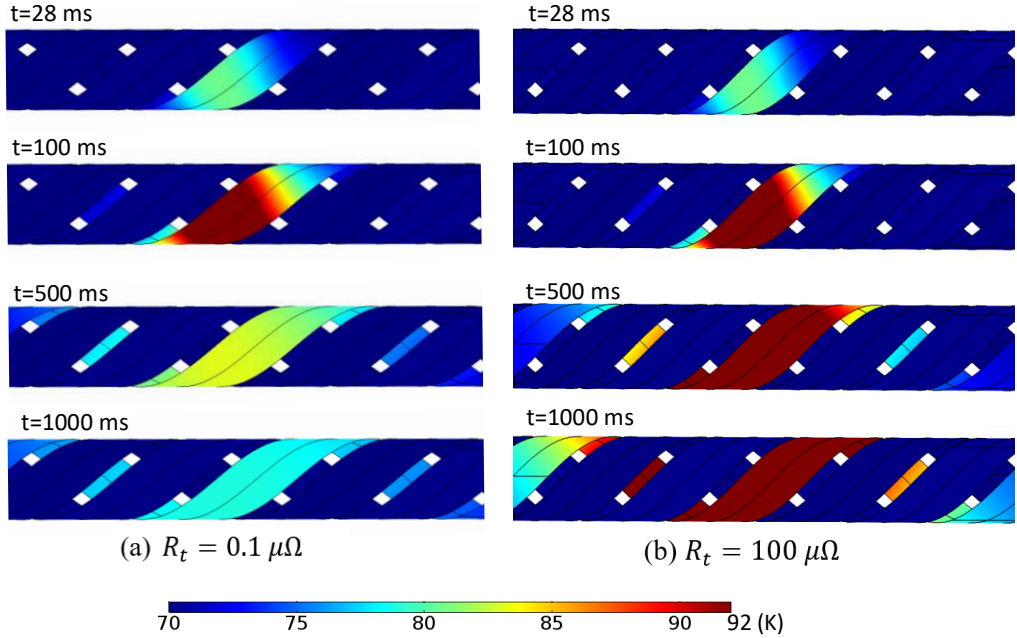

**Figure 10.** Normal zone propagation in the CORC cable when a heat pulse of 183 mJ is applied to Tape 1, total transport current is 360 A.

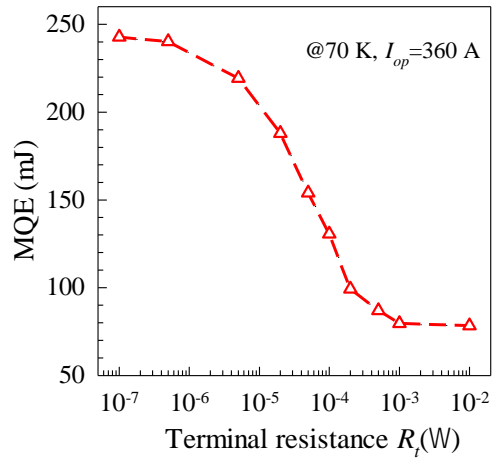

**Figure 11.** Dependence of minimum quench energy (MQE) on terminal contact resistance.

*3.2. Minimum quench energy*

The dependence of minimum quench energy (MQE) on terminal contact resistances has been shown in Figure 11. The result shows that the MQE drops rapidly with the increase of terminal contact resistance $R_t$ when the resistance is between $5\ \mu\Omega$ and $200\ \mu\Omega$. When the terminal contact resistance is higher than 100 $\mu\Omega$, the MQE reaches a minimum value. No further reduction of MQE is observed with further increase of the terminal contact resistance. When the terminal contact resistance is lower than 0.5 $\mu\Omega$, no further increase of MQE is observed with the decrease of the terminal contact resistance.

Figure 12 shows the current and maximum temperature of each tape during three typical local quenches. With the same heat pulse 243 mJ, the current redistributions among tapes are almost the same when $R_t = 0.1\ \mu\Omega$ and $R_t = 0.5\ \mu\Omega$, as shown in Figure 12($a_1$) and ($b_1$). The temperatures of the hot spot are similar, as shown in Figure 12($a_2$) and ($b_2$). When the terminal contact resistance is high, the current redistribution among tapes is very small. As shown in Figure 12($c_1$), with $R_t = 500\ \mu\Omega$, only 9 % of the transport current in Tape 1 is redistributed at the end of the heat pulse (when $t$=100 ms). The MQE at $R_t = 500\ \mu\Omega$ is 87 mJ, while the MQE at $R_t = \infty$ is 78 mJ. Therefore, when the terminal contact resistance is higher than 500 $\mu\Omega$, the thermal stability of CORC cable with multiple REBCO tapes connected in parallel has no obvious improvement in comparison to that of single REBCO tapes.

Results in Figure 12 also show that, with a low terminal contact resistance, the hot spot of one tape may induce an over-current quench on other tapes. As shown in Figure 12(a) and (b), the current redistribution can lead to an over-current condition on the tapes without hotspot, accompanied by a heat accumulation and a temperature rise. When the temperatures of Tape 2&3 increase, tape resistances start to increase, and the currents are redistributed again, flow back to tape 1. As shown in Figure12 ($a_1$) and ($b_1$), the current of Tape 1 after 3 s increases rapidly to 120 A and all three tapes are quenched. With high terminal contact resistance $R_t$, tape 2&3 are less affected by the hot spot in Tape 1. As shown in Figure 12 ($c_1$), although there is current redistribution to tape 2&3 from tape 1, there is no over-current quench on them. This is different from Figure 12 ($a_1$) and ($b_1$), where over-current quenches are generated in tape 2&3 due to current redistribution.

Whether there will be an overcurrent quench on all tapes depends on the operating current, depending on whether the rest of the tapes can take the redistributed current. Figure 13 shows a simulation with a total applied current of 300 A, which is 100 A per tape. The current of Tape 1 drops to nearly zero, and meanwhile the other tapes are still below critical current. Therefore, with low terminal contact resistances, it is impossible to induce a local quench on one tape when the other tapes can flow all the transport current below critical value.

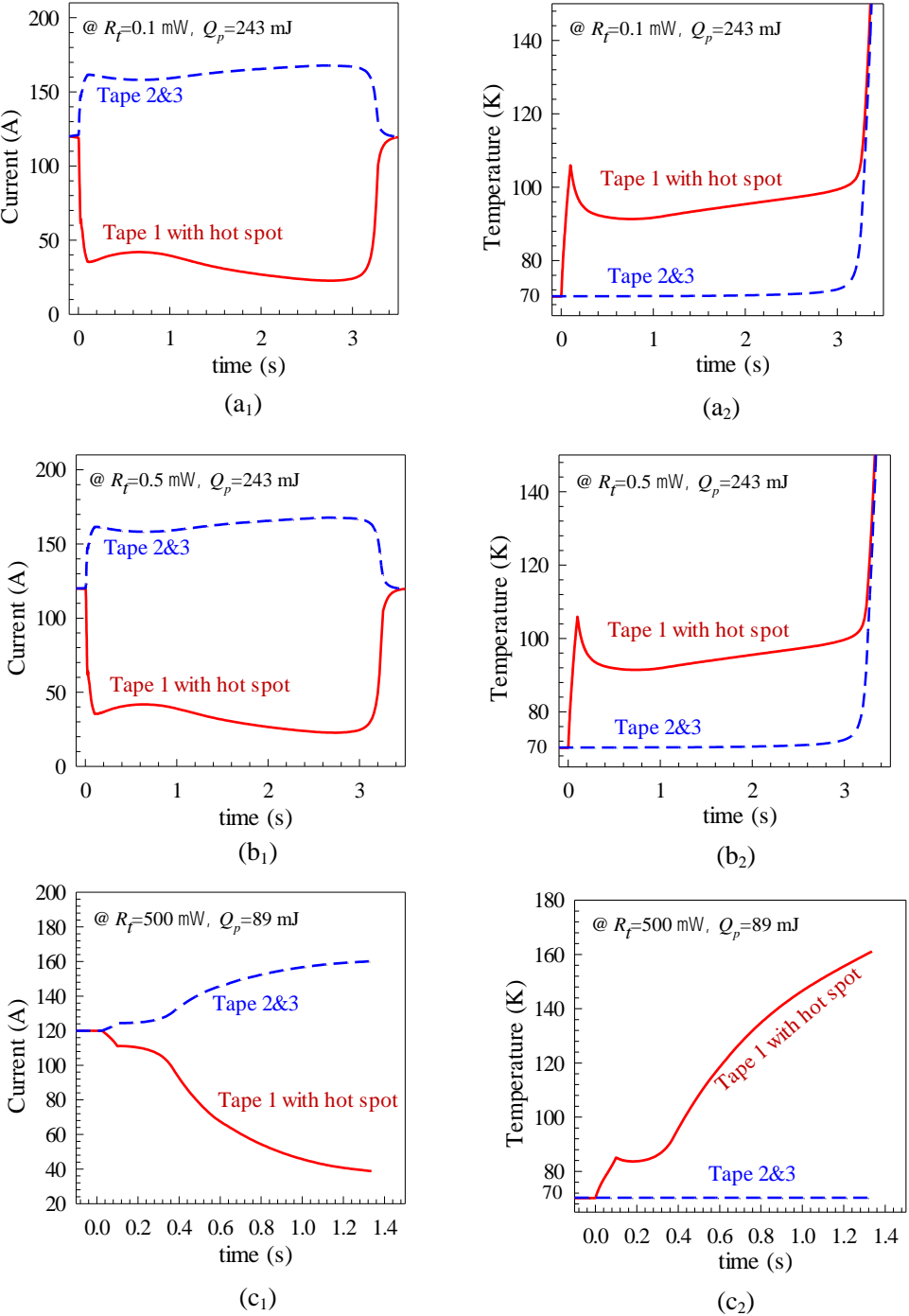

**Figure 12.** Current of each tape, maximum temperature of all the tapes during a hot spot induced quench operation; transport current is 120A per tape, heat pulse $Q_p$ is applied to Tape 1.

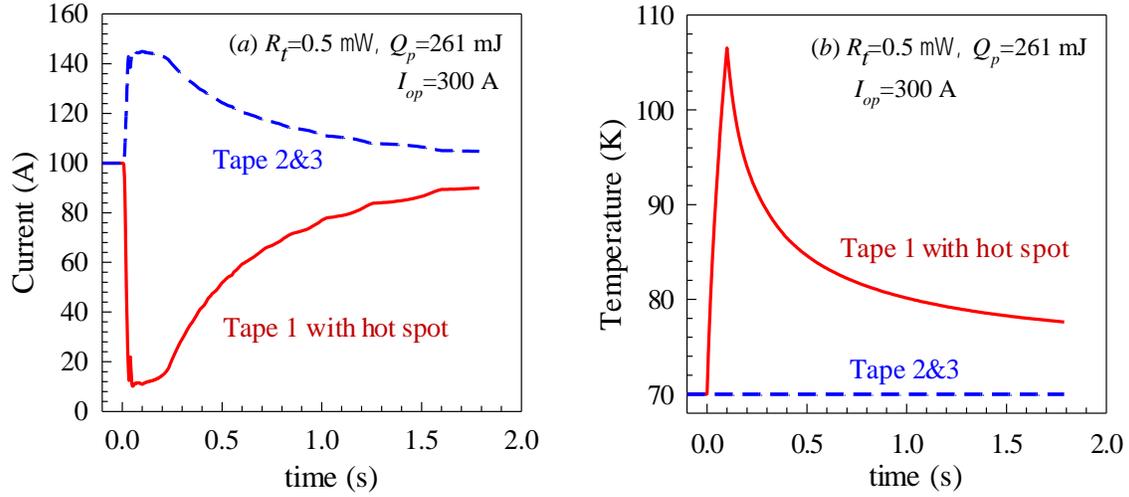

**Figure 13.** Current of each tape and maximum temperature of hot spot during a hot spot induced quench operation; transport current is 100A per tape, heat pulse $Q_p$ is applied to Tape 1.

## 5. Conclusions

In this paper, a 3D multi-physics model is developed based on the T-A formulation to study the physical process of hot spot induced quench of REBCO CORC cables. The multi-physics model coupled four modules: $T$-formulation model on shell, A-formulation model, a heat transfer model on shell, and an equivalent circuit model. The electromagnetic and thermal processes are studied using this model, and the influence of the terminal contact resistance on the quench behaviour is analysed and discussed.

Current redistribution occurs among REBCO tapes of CORC cable when a hot spot is induced on one tape. Some currents of the tape with hot spot will be forced out to other tapes without hot spot through terminal contact resistance. The quench risk of CORC cable can be minimized significantly by reducing the terminal contact resistance. With low terminal contact resistances, the current redistribution mainly depends on the critical current of the tape without hot spot. With high terminal resistances, the current redistribution mainly depends on the resistance.

The MQE increases rapidly with the reduction of terminal contact resistance $R_t$ when the resistance is in a middle range, which is between 5 $\mu\Omega$ and 200 $\mu\Omega$ in this study cases. When the terminal contact resistance is low enough, reducing the terminal contact resistance cannot further increase the MQE. When the terminal contact resistance is too high, the MQE drops to a ground value, which is similar to the MQE without terminal contacts. With a low terminal contact resistance, the hot spot of one tape may induce an overcurrent quench on the other tapes without hot spot. This will not happen in a cable with high terminal contact resistance. In this case, the tape with hot spot will quench and burn out before inducing another quench on other tapes.

The tape voltage and quench behaviour are significantly influenced by terminal contact resistances. Further work will be carried out in the future to develop quench detection and protection methods for CORC cables, as well as to experimentally validate the model proposed in this paper.

**Acknowledgments**

This project was supported by EPSRC grant EP/P002277/1. The authors would like to thank Dr. Anna Kario (Karlsruhe Institute of Technology) for sharing useful information about the geometry and structure of CORC cables. Dr Min Zhang would like to thank the support of Royal Academy of Engineering Research Fellowship. Dr Yawei Wang would like to thank the support of European Union's Horizon 2020 research and innovation programme under the Marie Sklodowska-Curie grant agreement No 799902.